\documentstyle[epsfig, twocolumn]{venice97}

\newcommand{\piH}{\pi_{\rm H}}
\newcommand{\piT}{\pi_{\rm T}}
\def\lessapprox{\,\raise 0.6ex\hbox{$<$}\kern -0.75em\lower 0.47ex
    \hbox{$\sim$}\,}

\begin{document}

%% Do not remove the following six lines:
\setlength{\parindent}{0pt}
\setlength{\parskip}{ 10pt plus 1pt minus 1pt}
\setlength{\hoffset}{-1.5truecm}
\setlength{\textwidth}{ 17.1truecm }
\setlength{\columnsep}{1truecm }
\setlength{\columnseprule}{0pt}
\setlength{\headheight}{12pt}
\setlength{\headsep}{20pt}
\pagestyle{veniceheadings}

%% Title - should be in capitals:
\title{SOME CONSIDERATIONS IN MAKING FULL USE OF THE HIPPARCOS CATALOGUE}

%% If the author list spans more than one line then the {\bf (bold
%% font)} command must be inserted for each line
\author{{\bf A.G.A. Brown$^1$, F. Arenou$^2$, F. van Leeuwen$^3$,
              L. Lindegren$^4$, X. Luri$^{2,5}$} \vspace{2mm} \\
$^1$Sterrewacht Leiden, Postbus 9513, 2300 RA Leiden, The Netherlands\\
$^2$Observatoire de Paris, Section d'Astrophysique, 5, Place Jules
Janssen, Meudon Cedex, 92195, France\\
$^3$Royal Greenwich Observatory, Madingley Road, Cambridge, CB3 0EZ,
United Kingdom\\
$^4$Lund Observatory, Box 43, S-22100, Lund, Sweden\\
$^5$Departament d'Astronomia i Meteorologia UB, Avda. Diagonal 647,
	E08028, Barcelona, Spain}

\maketitle

\begin{abstract}

This contribution is intended as a `rough guide' to the Hipparcos
Catalogue for the non-expert user. Some general aspects of the use of
astrometric data are discussed as well as Hipparcos-specific
applications. We discuss when and at what level one may expect systematic
errors to occur in the Hipparcos Catalogue. Next we discuss the
question of the interpretation of the measured parallaxes in terms of
distances and luminosities of stars. What are the biases one should be
aware of and how can these be corrected? When using the astrometric
data to study the statistics of stars one should take the full
covariance matrix of the errors on the astrometric parameters into
account. We explain how to do this and discuss the specific case of a
moving cluster. Finally, we address the question of the correlation of
astrometric parameters over a given region of the sky. At present the
Hipparcos Catalogue contains no identified systematic errors.
\vspace {5pt} \\
 
%% Do not remove the previous commands. Your abstract should 
%% end with \vspace {5pt} \\  

%% Please insert your keywords here.
Key~words: Space astrometry; Hipparcos; parallaxes; Luminosity
calibration; statistics.

\end{abstract}

\section{SYSTEMATIC ERRORS}\label{sec:systematics}

As opposed to `random error', the term `systematic error' is generally
understood to mean a statistical bias, i.e.\ that the error follows a
distribution with mean value (or some other measure of location)
different from zero. The application of this statistical concept to
the Hipparcos Catalogue is far from trivial.  To begin with, the
Hipparcos Catalogue is unique and cannot be repeated.  Is it then
meaningful to speak of the bias of an individual data item in the
catalogue?  It probably is, as much as it is meaningful to speak of
the standard error of a single datum: both depend on the notion that
the observed value is `drawn' from a population with a definite
statistical distribution.  In practice, however, the separation of
random and systematic errors requires averaging, and the only
averaging possible in our case is with respect to a sample of {\em
different\/} stars.  It is then necessary to assume that the stars in
this sample share similar statistical properties.

%A systematic error need not be a constant error. More likely it will
%depend on factors such as position on the sky, colour or magnitude, or
%the detailed scanning geometry of a given object. If many such factors
%are at play the combined effect may be practically indistinguishable
%from a random error.

Apart from these formal difficulties, the analysis of the Hipparcos
Catalogue with respect to systematics faces a very severe practical
problem. Systematic errors can generally only be revealed through
comparison with independent data of at least similar quality.  Very
few such data exist and the tests that have been performed on the
Hipparcos data are therefore limited in scope and precision.  The
results of several comparisons are summarized below; for a full
description see Chapters~18 to 22 in Volume~3 of the Hipparcos
Catalogue (ESA 1997).

The published catalogue is essentially the mean of the two separate
reductions performed by the FAST and NDAC consortia.  While a
comparison of the two reductions does not prove anything about the
systematic errors of the final catalogue, it gives considerable
insight into the properties of the errors. Thus we may perhaps take
the systematic FAST/NDAC differences (see Volume~3, Chapter~16) as an
indication of what can be expected for the systematic errors in the
Hipparcos Catalogue.

\subsection{Position and Proper Motion}   

The positions and proper motions in the Hipparcos Catalogue formally
refer to ICRS, the International Celestial Reference System replacing
(although closely coinciding with) the `equinox 2000' system. ICRS is
defined by means of extra-galactic radio sources and great care was
taken to link the Hipparcos Catalogue to this extra-galactic system
(\cite{kova97}).  The final uncertainty of the link corresponds to an
orientation error of $\pm 0.6$~milliarcsec (mas) for the system of
positions at the epoch J1991.25, and to an error of $\pm
0.25$~mas~yr$^{-1}$ for the global rotation of the proper motion
system.  For the epoch J2000 the uncertainty in the orientation of the
Hipparcos positions with respect to ICRS will increase to
$[0.6^2+(8.75\times0.25)^2]^{1/2}\simeq \pm 2.3$~mas.  The difference
between the Hipparcos positions and proper motions (known as the
Hipparcos reference frame) and the ICRS may be regarded as a
systematic error of the catalogue.  The uncertainties of the
extra-galactic link quoted above are not included in the standard
errors of the positions and proper motions of Hipparcos objects as
given in the catalogue.

Other systematic errors in the positions and proper motions correspond
to a distortion of the Hipparcos reference frame, and consequently
affect e.g.\ the calculated angle between objects.  Practically the
only significant external check was achieved by means of the 12 radio
stars observed by VLBI, yielding rms residuals of $1.7$~mas in
position (epoch J1991.25) and $0.8$~mas~yr$^{-1}$ in proper motion.
These are consistent with the formal standard errors (taking into
account the known structure of two of the objects), indicating that
the distortions of the Hipparcos reference frame are less than 1~mas
and $0.5$~mas~yr$^{-1}$, respectively.  Differences between the NDAC
and FAST reductions suggest errors of a similar size on a very local
scale (few degrees).  Large-scale systematic differences are
considerably smaller, e.g.\ $<0.1$~mas or mas~yr$^{-1}$ on a scale of
$90^\circ$.
 
\subsection{Parallax}   

A global zero-point error in the Hipparcos parallaxes could in
principle be produced by a specific harmonic of a systematic variation
of the instrument with respect to the solar aspect angle. Such
possible variations were guarded against in the satellite thermal
design, and were carefully investigated during data reduction, leading
to the conclusion that any global effect of this nature is probably
less than $0.1$~mas.  {\em A~priori\/} we thus expect the Hipparcos
parallaxes to be absolute.

A comparison of Hipparcos parallaxes with the best ground-based
optical parallaxes (88 stars; from the USNO 61-inch reflector) gives a
median difference of $+0.2\pm 0.35$~mas, suggesting the absence of
systematic differences between the two techniques.  Parallaxes of
radio stars obtained by VLBI are also in very good agreement with
Hipparcos.  Comparison with other ground-based parallax programmes
(see \cite{GCTP}) shows systematic differences of up to several
milliarcsec, especially for the southern sky; part of this may be
related to the transformation from relative to absolute parallax in
the ground-based programmes.

Using the photometric distances of open clusters more than 200~pc
away, a parallax zero-point error of $+0.04 \pm 0.06$~mas was derived.
For a sample of 467 field stars with $uvby\beta$ photometry, the
statistical method of \cite*{aren95} gave a zero-point error of $-0.05
\pm 0.05$~mas. From these comparisons the global zero-point error of
the Hipparcos parallaxes is considered to be smaller than
$0.1$~mas. However, note that in general very red stars may exhibit
various problems, including a possible zero point error. For details
we refer to Chapters~20 and 21 of Volume~3 of the Hipparcos Catalogue.

\subsection{Photometry}

Although Hipparcos was not specifically designed for accurate
photometry, the all-sky photometric survey in the $Hp$, $B_T$ and
$V_T$ bands provides a data base of unprecedented homogeneity.  No
significant systematic errors are expected as a function of position.
However, small non-linearities of the magnitude scales, partly due to
a saturation effect in the Hipparcos measurements, are found through
comparison with ground-based Johnson and Geneva photometry: for the
$Hp$ scale, a mean slope of $-0.0017$~mag~mag$^{-1}$ in the range
$V=3$ to 9~mag and departures up to $0.04$~mag around $V=0$; for $B_T$
and $V_T$ systematic deviations occur instead at the faint end as a
result of statistical biases. For details refer to Chapter~21 of
Volume~3 of the Hipparcos Catalogue.

The temporal stability of the magnitude scales is generally superb,
permitting the detection of variability at the level of a few
hundredths of a magnitude.  However, radiation darkening of the optics
caused a significant variation of the instrument passbands which had
to be taken out in the photometric reductions. If the reduction was
made with an erroneous $V-I$ colour index, this may have produced a
spurious trend in the $Hp$ magnitudes. The value of $V-I$ used for the
reductions and a procedure for correcting any such trend if an
improved $V-I$ becomes available, are given in the Hipparcos Catalogue
(Volume~3, Chapter~14).

\subsection{Outliers and External Accuracy}   

Related to the statistical distribution of the errors in the catalogue
is the question of outliers (i.e.\ errors exceeding what can
reasonably be expected of a Gaussian distribution) and external
accuracy (i.e.\ the actual standard deviation of errors compared with
the stated formal standard errors).  A very small number of gross
errors in position may exist, especially among the double-star
components, as caused by grid-step errors ($>0.5$~arcsec).  The proper
motions and parallaxes are generally less susceptible to this kind of
error.  For the proper motions one should however be aware that
unresolved duplicity (astrometric binaries) may produce significant
differences with respect to ground-based values (\cite{lind97};
\cite{wiel97}).  For the parallaxes a similar effect can occur in the
very rare case of an unrecognized binary with a period of about one
year.  In the epoch photometry, outliers occasionally occur, caused
by satellite attitude errors (giving reduced flux) or parasitic stars
from the complementary field of view (giving increased flux).

A meaningful check of the external accuracy has only been possible in
the case of the parallaxes, through comparisons with photometric
distances.  These indicate that the external standard errors are about
$1.05 \pm 0.05$ times larger than the standard errors given in the
catalogue, at least for the brighter stars ($V<9$~mag). From the
general method by which the parallaxes were computed, it is reasonable
to assume that the same factor applies to the standard errors in
position and proper motion of single stars.  The situation is much
more complex for resolved double and multiple stars, but as a general
rule it is believed that the errors are not underestimated by more
than a factor 1.2.

\section{CORRECT USE OF TRIGONOMETRIC PARALLAXES}

Notwithstanding the unprecedented quality of the Hipparcos data, the
correctness of the {\em astrophysical\/} results is not assured, as
the estimation of stellar distances, absolute magnitudes and other
physical quantities from trigonometric parallaxes is not a trivial
process. The statistical properties of the relationships involved and
the effects of sample selection hide several pitfalls that, if not
avoided, lead to biased estimates.

We assume for this discussion that the Hipparcos parallaxes are
unbiased, in the sense that their systematic errors are small compared
to their random errors (see Section~1). Nevertheless, biases in the
derived results may occur if an improper analysis of the data is
done. In this section we present a brief review of the statistical
properties of trigonometric parallaxes and derived quantities, as well
as the effects of sample truncation(s). References given at the end of
this section may be consulted for the work done up to now on avoiding
the various biases and making full use of the trigonometric
parallaxes.

%++++++++++++++++++++++++++++++++++++++++++++++++++++
\subsection {Selection Biases} \label{sec:artefacts}
%++++++++++++++++++++++++++++++++++++++++++++++++++++
A well-known selection bias is the \cite*{ma36} bias. In this case, a
set of non-biased apparent magnitudes leads to a biased mean absolute
magnitude due to the combination of the apparent magnitude limit of
the sample and the intrinsic dispersion of absolute magnitudes (e.g.,
\cite{LU93}). In statistical terms: the selection criteria make the
mean absolute magnitude of the sample non-representative of that of
the underlying parent population, thus introducing a bias, as faint
stars are underrepresented.  The use of the parallaxes of a truncated
sample without caution may lead to similar biases in the derived
results (see poster 3.45 for some common examples).

Let us assume, for instance, that we want to check the systematic
difference between Hipparcos ($\piH$) and Tycho parallaxes ($\piT$)
and that for this purpose we select a sample containing only stars
with $\piH<1$ mas.  Computing the median difference $\piT-\piH$ on
this sample results in $0.28\pm0.01$ mas, which suggests a significant
systematic error in either the Tycho or Hipparcos parallaxes. However,
this is only a selection bias due to the combination of the criterion
$\piH<1$ mas, the non-uniformity of the parallax distribution and the
random errors in $\piH$ and $\piT$. Indeed, the median difference
$\piT-\piH$ without truncating the parallax distribution is not
significantly different from zero. This example clearly illustrates
how a truncation in the observed parallax distribution can introduce a
bias in the sample so that, even if the individual parallaxes are not
biased, the computed mean is biased.  This example is based on
truncation of the observed parallaxes, but the same argument applies
to a selection based on the relative parallax errors
$\sigma_{\piH}/\piH$.

Another type of bias is caused by an indirect truncation of the
parallaxes. For instance, suppose that the spatial velocities of a
given sample are computed and stars with high spatial velocity are
selected. This will select stars with a truly high velocity but also
stars with an overestimated distance ($\piH \ll \pi$): the estimated
distances of objects in this subsample will be biased in the
mean. Consequently its estimated mean absolute magnitude will be too
bright.

%++++++++++++++++++++++++++++++++++++++++++++++++
\subsection {Biased Estimates}\label{sec:biases}
%++++++++++++++++++++++++++++++++++++++++++++++++
Several quantities, such as the stellar distance or the absolute
magnitude, have a non-linear dependence, $h(\pi)$, on the parallax. In
this case, the expectation value of the function, $E[h(\piH)]$, is in
general different from $h(\pi)$, even if the individual Hipparcos
parallaxes are unbiased, i.e.\ if $E[\piH]\approx\pi$. In other words,
$1/\piH$ is a biased estimate of the star's true distance, and
$m+5\log\piH+5$ is a biased estimate of its absolute magnitude:
$E[{1/\piH}] \neq {1/\pi}$ and $E[m+5\log\piH+5] \neq m+5\log\pi+5$.

%______________________________________________________________________________
\begin{figure}[!ht]
  \begin{center}
    \leavevmode
    \centerline{\epsfig{file=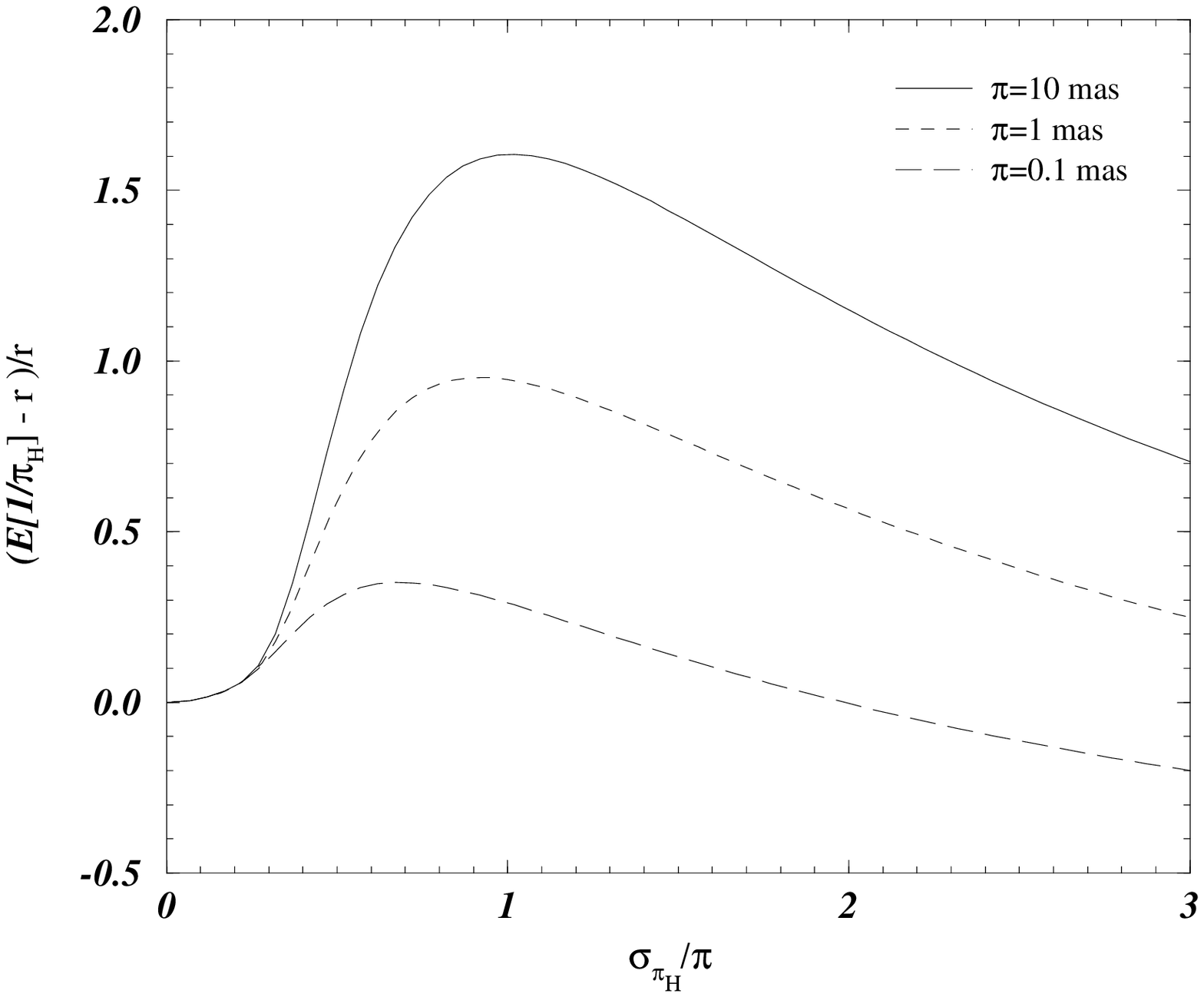,width=7.5cm}}
    \centerline{\epsfig{file=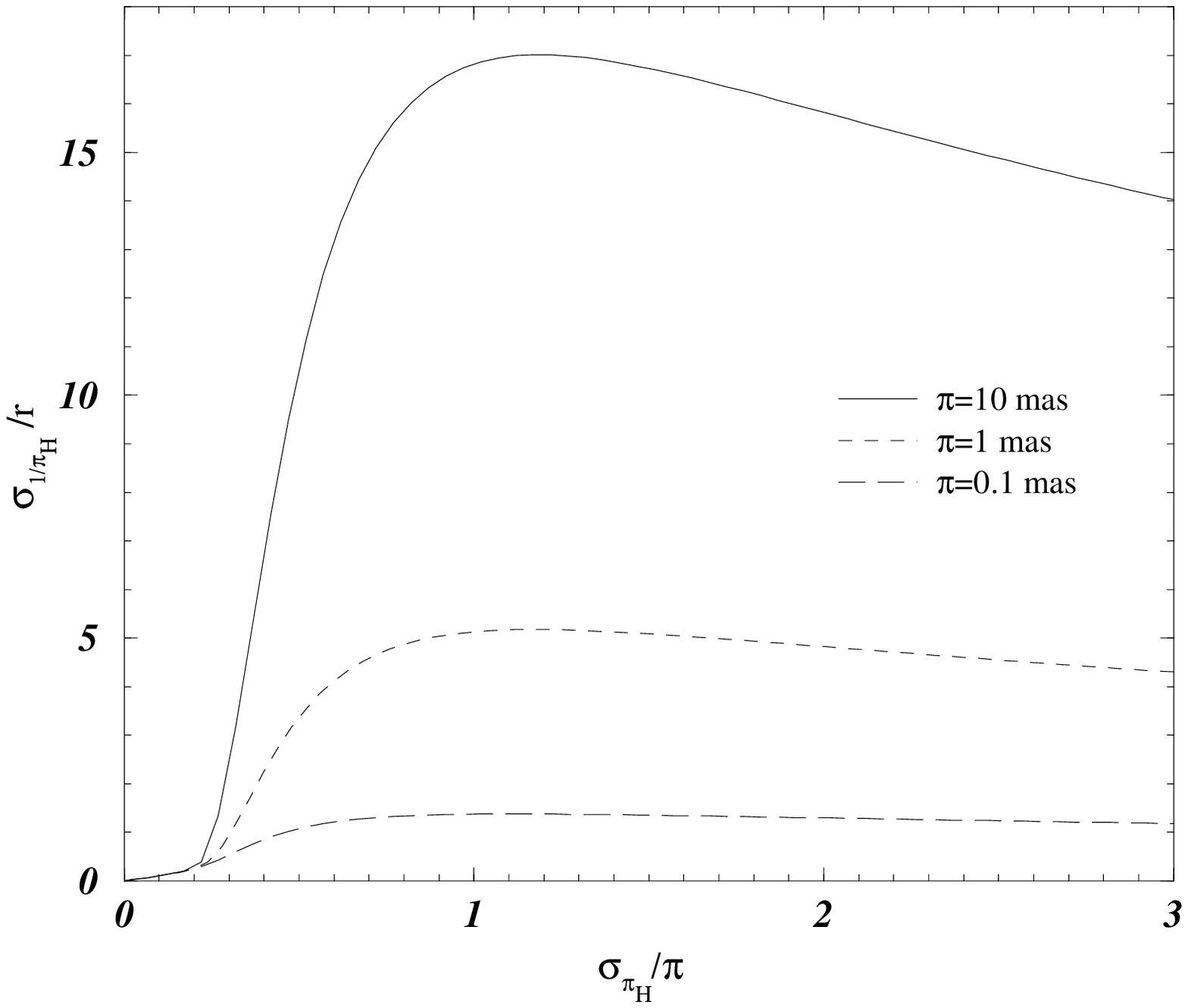,width=7.5cm}}
    \end{center}
    \caption{\em Relative bias (top) and relative precision (bottom)
    of computed distance as a function of the ratio of the parallax
    observational error to the true parallax.}
    \label{fig:biasig-dist}
\end{figure}
%______________________________________________________________________________

%______________________________________________________________________________
\begin{figure}[!ht]
  \begin{center}
    \leavevmode
    \centerline{\epsfig{file=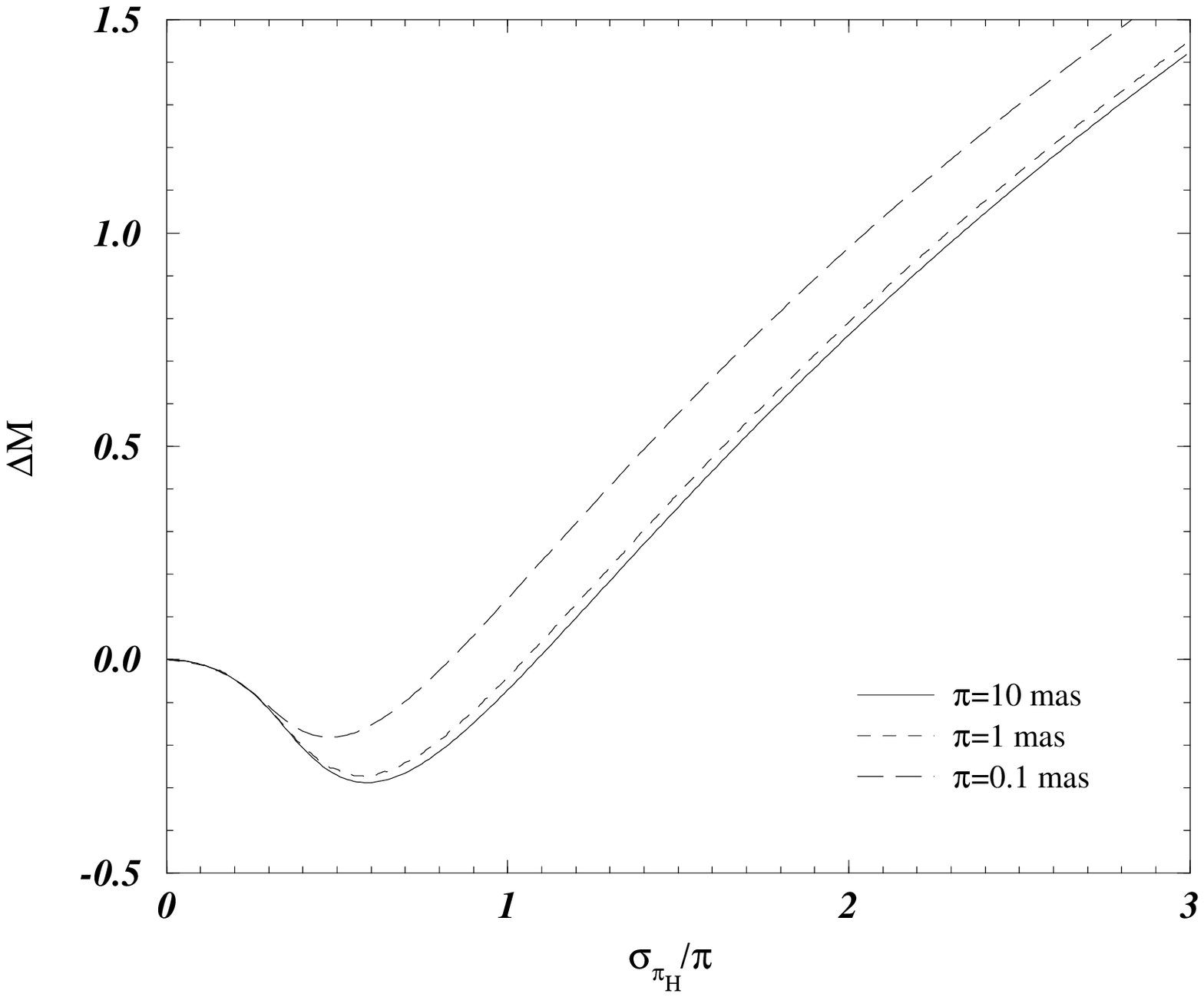,width=7.5cm}}
    \centerline{\epsfig{file=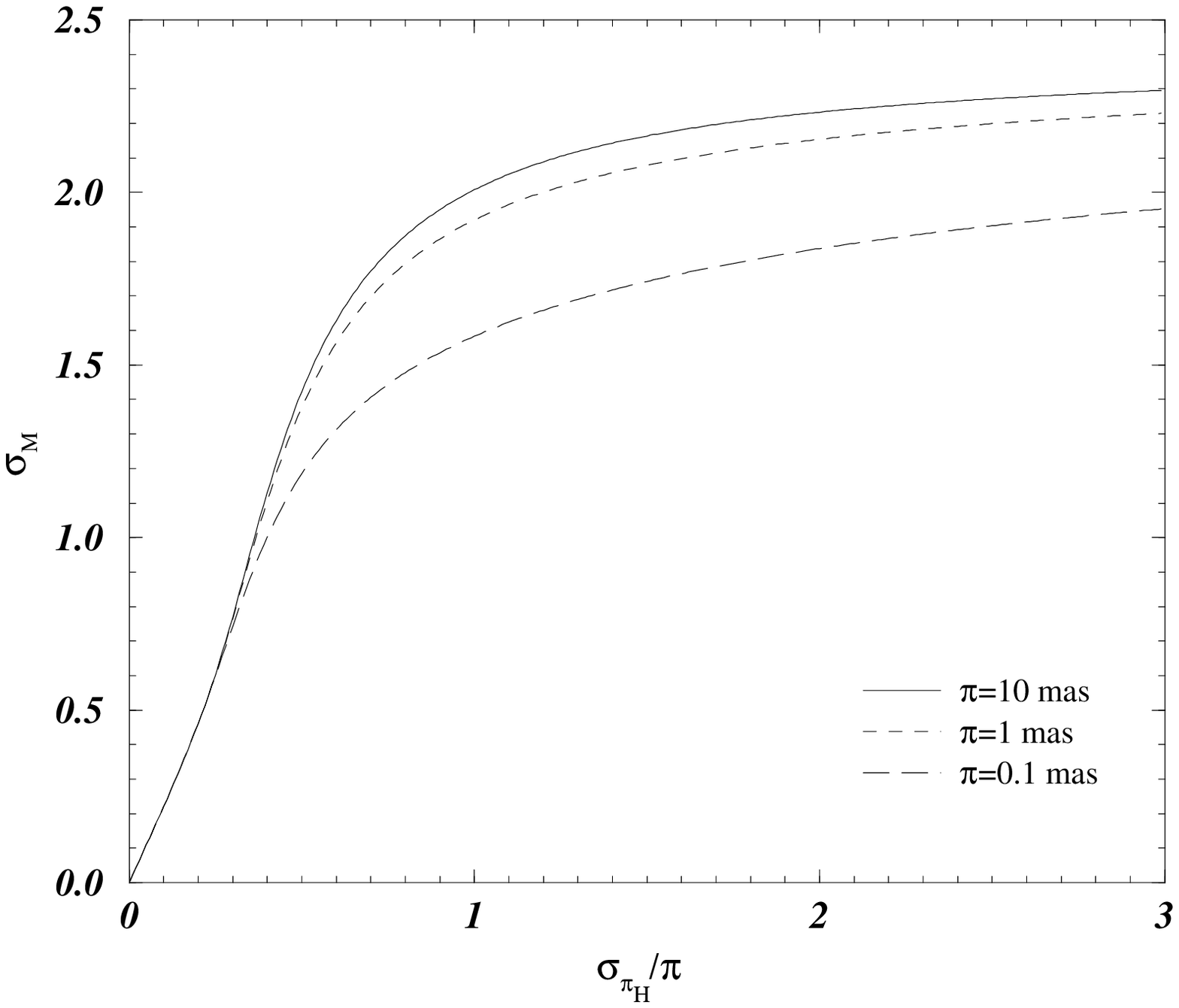,width=7.5cm}}
  \end{center}
  \caption{\em Bias (top) and precision (bottom) of computed absolute
  magnitude as a function of $\sigma_{\pi_{\rm H}}/\pi$.}
  \label{fig:biasig-absm}
\end{figure}
%______________________________________________________________________________
%

To study this problem we will exclude negative and zero parallaxes, as
will many users of the Hipparcos Catalogue. The lower bound used in
the following calculations is $0.01$~mas, which is the smallest
non-zero value of parallax that can be found in the catalogue.  In
Figure~\ref{fig:biasig-dist} the value of the relative bias
$(E[1/\piH]-r)/r$ is shown as a function of the ratio of the
observational error to the {\it true} parallax, for several values of
this parallax. The computed distance may be overestimated by more than
100~per cent when the relative error in $\pi$ is 100~per cent. Note
that if negative parallaxes are not rejected, the bias, although
reduced, is still present (\cite{SE96}).

Since zero parallaxes have been rejected, the variance of $1/\piH$ is
not infinite and may be computed. The calculation is depicted in
Figure~\ref{fig:biasig-dist}. This is to be compared to the usual
first order approximation ${\sigma_r/r}\approx{\sigma_{\pi}/\pi}$,
which is valid to within $\sim25$~per cent approximately up to a 20~per
cent relative error.

Figure~\ref{fig:biasig-dist} shows that both bias and variance are
negligible for relative errors better than about 10~per cent. For
parallaxes with a higher relative error, one could naively hope to
correct the computed distance from the bias shown above, but this is
not possible because the bias is a function of $\sigma_{\piH}/\pi$,
where the {\em real\/} parallax $\pi$ unknown. What is available is
{\em not\/} the real relative parallax error but the observed one
$\sigma_{\piH}/\piH$. Note that Figure~\ref{fig:biasig-dist} {\it
also} indicates the relative bias of $\sigma_{\piH}/\piH$ as an
estimate of $\sigma_{\piH}/\pi$ and its relative precision.  Given the
uncertainty on the true relative error, a bias correction is simply
not feasible. On the other hand, such a correction would only have a
statistical meaning when applied to a sample, but would be
questionable on an individual basis.

The situation is the same when considering absolute magnitudes
(Figure~\ref{fig:biasig-absm}). The absolute magnitudes computed from
observed parallaxes are almost unbiased for small relative errors
($\sigma_{\pi_{\rm H}}/\pi\lessapprox 0.1$), but are on the average
$0.2$ mag too bright when the relative parallax error is about 50~per
cent, and $0.6$ mag too faint for a 200~per cent relative
error. Again, the correction for these biases would in principle
require knowledge of the {\em true\/} parallaxes.

%+++++++++++++++++++++++++++++++++++++++++++++++++++++++++++++++++++++++++++++
\subsection {How to use Hipparcos Trigonometric Parallaxes}\label{sec:whatodo}
%+++++++++++++++++++++++++++++++++++++++++++++++++++++++++++++++++++++++++++++
Various methods to use astrometric data with minimal biases have been
proposed in the past and are summarized below, together with their
positive and negative aspects:
\begin{itemize}
  \item Only stars with the best relative errors are kept.  Keeping
        only stars with ${\sigma_{\piH}/\piH}<10~$per cent means that
        more than 20\,000 stars are still available. However, due to
        the implicit truncation of the parallax, a bias should still
        be expected.

  \item \cite*{SE96} propose another estimator of distance, and
        absolute magnitude, based on a transformation of the observed
        parallax.  Although the bias and variance of the new estimates
        are reduced, their physical meaning is questionable.

  \item Models using all available information (photometry, position,
        proper motion) can be built in order to derive unbiased and
        precise estimates of physical data of interest: absolute
        magnitude, distance, kinematics (\cite{RC91}, \cite{LU96},
        \cite{aren95}). The drawback is, of course, that the estimates
        found are model-dependent.
        
  \item Finally, a recent approach using Hipparcos intermediate data
  	has been proposed by \cite*{vleeu97}, for the
  	calibration of absolute magnitudes. Using no parametric model
  	and all the available data, there remains however a correction
  	to be done for magnitude-limited samples.
\end{itemize}
Summarizing, one can easily calculate the expected biases for a {\em
given\/} true parallax. However, one only has the observed values, so
the correction will depend on what kind of assumption one makes
concerning the true values. In other words, the {\em distribution\/}
of the true parallaxes has to be known, and this is an astrophysical
question, not a statistical one! Hence, one cannot solve this problem
just by statistics, but needs also some kind of modeling of the objects or
sample under study. The reader is strongly encouraged to perform a
detailed analysis of this sort {\em for each specific case\/} in order
to obtain a correct estimation of any parameter of a star or a sample
of stars using trigonometric parallaxes. This means in particular that
one should neither ignore the possible biases nor apply blindly
`Malmquist' or `Lutz-Kelker' corrections (\cite{LK73}).

\section{USE OF THE COVARIANCE MATRIX}
\label{sec:covar}

One of the unique features of the Hipparcos Catalogue is that not only
the standard errors of the five astrometric parameters are provided
but also their correlation coefficients. This allows the user to make
full use of the information contained in the astrometric
parameters. In the following we demonstrate briefly the use of the
covariance matrix and we show the importance of using the matrix with
a worked example. Here we concentrate on using the covariance matrix
when interpreting the statistics of a particular data set. The
covariance matrix is also necessary if one is interested in
propagating the positions, proper motions and the corresponding
standard errors within ICRS to an epoch different from the epoch --
J$1991.25$ -- of the Hipparcos Catalogue. Propagation routines in C
and Fortran are provided in the catalogue. For more information on the
covariance matrix in relation to the astrometric parameters please
refer to Sections $1.2$ and $1.5$ in Volume~1: Part~1 of the Hipparcos
and Tycho Catalogues (\cite{ESA97}).

If $\bf x$ is an observed vector with covariance matrix $\bf C_{\bf
x}$ then the confidence region around $\bf x$ is given by: $c={\bf
x}'{\bf C}_{\bf x}^{-1}{\bf x}$, where the prime denotes matrix
transposition. The distribution of $c$ is described by a $\chi^2_\nu$
probability distribution, where $\nu$, the number of degrees of
freedom, is equal to the dimension of $\bf x$. In the one-dimensional
case this reduces to the well-known Gaussian distribution, where $c=9$
corresponds to `$3\sigma$', the $99.73$~per cent confidence level. For
other values of $\nu$ the value of $c$ will be higher for the same
confidence level. It is $11.8$ for $\nu=2$ and $14.2$ for $\nu=3$.
Note that the distribution of the errors around $\bf x$ is described
by a multi-dimensional Gaussian and the equation above describes a
confidence `ellipsoid' around $\bf x$.

If the vector $\bf y$ is derived from $\bf x$ via some transformation
${\bf f}({\bf x})$, the covariance matrix of $\bf y$ is: ${\bf C}_{\bf
y}={\bf JC_{\bf x}J}'$. Here $\bf J$ is the Jacobian matrix of the
transformation from $\bf x$ to $\bf y$: $\lbrack{\bf
J}\rbrack_{ij}=\partial f_i/\partial x_j$. Thus one can calculate the
covariance matrix of any set of variables derived from the observed
astrometric parameters.

We now turn to the example of space velocities for cluster stars,
specifically the Hyades. For the full details we refer the reader to
\cite*{Perryman97}. When deriving space velocities for cluster stars
we make use of the observed vector
$(\pi,\mu_{\alpha*},\mu_\delta,V_{\rm R})$, where $V_{\rm R}$ is the
radial velocity. This vector is transformed to a space velocity,
implicitly invoking a transformation to $(V_{\alpha*},V_\delta,V_{\rm
R})$ ($V_{\alpha*}=\mu_{\alpha*}A_v/\pi, V_\delta=\mu_\delta A_v/\pi$,
$A_v=4.74047...$~km yr s$^{-1}$). To emphasize that using the
covariance matrix is important even if the observed parameters are
uncorrelated we shall proceed on the assumption that the astrometric
errors are uncorrelated. Then the transformation of the observables to
the vector $(\pi,V_{\alpha*},V_\delta,V_{\rm R})$ yields the
covariance matrix:
\begin{equation}
  \left( \begin{array}{cc}
    {\bf S} & \emptyset\\[+2pt]
    \emptyset & \sigma_{V_{\rm R}}^2
  \end{array} \right) \,,
  \label{eq:eqcovar}
\end{equation}
With $a=A_v/\pi^2$, $\bf S$ is given by:
\[ \arraycolsep 4pt
   \footnotesize
   \left( \begin{array}{ccc}
\sigma_\pi^2 & -\mu_{\alpha*}a\sigma_\pi^2 &
               -\mu_\delta a\sigma_\pi^2 \\[+2pt]
-\mu_{\alpha*}a\sigma_\pi^2 &
a^2 \mu_{\alpha*}^2 \sigma_\pi^2 +
      A_v a \sigma_{\mu_{\alpha*}}^2 &
a^2 \mu_{\alpha*} \mu_\delta \sigma_\pi^2 \\[+2pt]
-\mu_\delta a\sigma_\pi^2 &
a^2 \mu_{\alpha*} \mu_\delta \sigma_\pi^2 &
a^2 \mu_{\delta}^2 \sigma_\pi^2 +
      A_v a \sigma_{\mu_\delta}^2 
\end{array} \right) \,. \]
Hence, even in the absence of correlations between astrometric errors,
the parallaxes and velocity components $V_{\alpha*}$ and $V_\delta$
will in general be correlated. Moreover, because of the position of
the convergent point of the Hyades with respect to the cluster centre,
$\mu_{\alpha*}$ is positive and $\mu_\delta$ is negative for most
cluster members, and hence the product ${\mu_{\alpha*}}\mu_\delta$ is
negative. Thus for the Hyades the correlated errors will lead to
systematic behaviour of the uncertainties in the sample as a
whole. These systematics will be transferred to the space velocities.

%%%%%%%%%%%%%%%%%%%%%%%%%%
%
% Velocity vectors figure
%
%%%%%%%%%%%%%%%%%%%%%%%%%%
\begin{figure}[!ht]
  \begin{center}
  \leavevmode
  \centerline{\epsfig{file=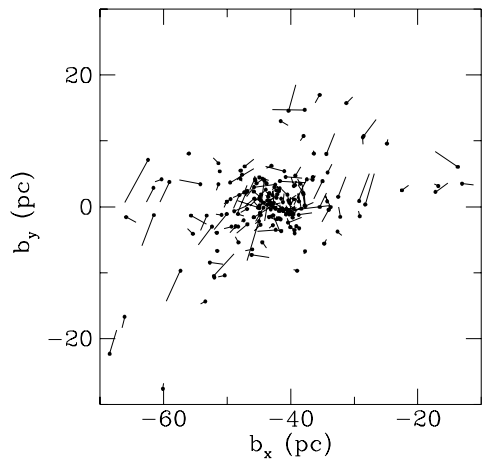,width=6.5cm}}
  \end{center}
  \caption{\em Projected velocities as a function of position for
  Hyades members. The residuals are given as vectors with respect to
  the mean velocity of the cluster in Galactic coordinates}
  \label{fig:velvec}
\end{figure}

Figure~\ref{fig:velvec} shows the velocity of the Hyades members with
respect to the mean cluster motion plotted as vectors on the Galactic
$x$-$y$ plane. One immediately picks out a systematic motion,
suggestive of rotation or shear. However, what one sees is a
correlation between the velocity residuals (magnitude and direction)
and the distances (parallaxes) of the stars. This can be understood as
follows. The difference between the {\em observed\/} and {\em true\/}
stellar parallaxes ($\Delta\pi=\pi_{\rm obs}-\pi_{\rm true}$) is not
correlated with the true parallaxes. However, because all Hyades
members have similar parallaxes, adding $\Delta\pi$ to $\pi_{\rm
true}$ implies that, on average, the stars with positive $\Delta\pi$
will have the largest {\em observed\/} parallaxes (and vice versa for
the stars with negative $\Delta\pi$). So the sign of the parallax
error is correlated with the {\em observed\/} parallax. The correlation
between $\Delta\pi$ and $V_{\alpha*}$ and $V_\delta$, discussed above,
will then lead to a correlation between the observed distances of the
stars and the velocity residuals.

Figure~1 in \cite*{Brown97} shows how one can explain both the total
spread and the correlations between velocity components by considering
the covariance matrix of the observations.  Hence, in the case of the
Hyades both the overall distribution of the velocity residuals, as
well as the correlation of the direction of the residuals with spatial
position (the features in Figure~\ref{fig:velvec}), can be fully
attributed to observational errors.

We stress here that ignoring the covariance matrix can easily lead to
false interpretation of, for example, kinematic data. For cases other
than the Hyades the way in which the features due to correlated errors
enter may differ. It is important to carry out this kind of analysis
and consider the implications for each case individually.

\section{CORRELATION OF ASTROMETRIC PARAMETERS ON THE SKY}

The Hipparcos data for stars concentrated in a small area of the sky
have been derived from partly correlated observations (see Volume~3,
Chapter~17 of \cite{ESA97}). This means that proper motions and
parallaxes of stars in open clusters or in the Magellanic Clouds, for
example, cannot be interpreted as fully independent observations. For
instance, the parallax errors of stars within a small ($<2^\circ$)
area of the sky in general have a positive statistical correlation
($\rho>0$) because the stars were observed in more or less the same
scans and part of their parallax errors derive from abscissa errors
which were constant within each scan.  Averaging the parallax errors
of $n$ stars in such an area will not quite produce the expected
improvement by $n^{-1/2}$; in fact the error approaches (in principle)
a certain limiting value as $n$ is increased indefinitely, exactly as
in the presence of a systematic error. Estimates suggest that the
average of $n$ stars improves as $n^{-0.35}$ for stars separated by
less than about 2$^\circ$.

The data from which the astrometric parameters have been derived have
been preserved in the `Hipparcos Intermediate Astrometric Data' file
on Disc~5 of the ASCII CD-rom set. Using those data, the correlations
can be taken into account. Full details on this procedure, as well as
more background information on the correlations in the Hipparcos
observations are given by \cite*{vleeu97}. The intermediate data also
allow solutions in which information on astrometric parameters is
linked within a selection of stars, such as stars in a cluster or
stars sharing the same luminosity characteristics. In these cases, the
individual parallax and/or proper motion solutions for the individual
stars are replaced by the solution of a few common parameters for all
stars involved, describing for example the parallax as a constant
value (for a cluster) or as a function of photometric and
spectroscopic parameters. For a specific example, where this is
applied to the Pleiades, see \cite*{leeuw97} and \cite*{mermio97}.

\section{SELECTION AND COMPLETENESS}

Finally, we want to end by emphasizing that in {\em any\/}
quantitative study of a sample of stars it is {\em essential\/} to
take {\em selection effects and completeness\/} into
account. Unfortunately in the specific case of Hipparcos it is not at
all trivial do so, even though a specific effort has been made to
carry out part of the Hipparcos mission as a survey which is roughly
complete to $V\sim7$--8~mag. We will not discuss this issue here but
refer the reader to \cite*{turon92} and Volume II of \cite*{ESA89},
specifically Chapters~7 and 8. These references describe the details
of the construction of the Hipparcos Input Catalogue and also go into
the details of the catalogue completeness.

\end{document}